\documentclass[12pt]{article}

\usepackage{graphicx}% Include figure files
\usepackage{color}
\usepackage[colorlinks=true, urlcolor=blue, linkcolor=blue, 
citecolor=blue]{hyperref}
\usepackage{cite}

\newcommand{\beq}{\begin{equation}}
\newcommand{\eeq}{\end{equation}}
\newcommand{\beqa}{\begin{eqnarray}}
\newcommand{\eeqa}{\end{eqnarray}}
\newcommand{\beqar}{\begin{eqnarray*}}
\newcommand{\eeqar}{\end{eqnarray*}}

\begin{tiny}

\end{tiny} %{\vskip-2ex$_{#1}%$\label{#1}}

\begin{document}
\thispagestyle{empty}
$\,$

\vspace{32pt}

\begin{center}

\textbf{\Large High energy neutrinos from the Sun}

\vspace{50pt}
Manuel Masip 
\vspace{16pt}

\textit{CAFPE and Departamento de F{\'\i}sica Te\'orica y del Cosmos}\\
\textit{Universidad de Granada, E-18071 Granada, Spain}\\
\vspace{16pt}

\texttt{masip@ugr.es}

\end{center}

\vspace{30pt}

\date{\today}% It is always \today, today,

\begin{abstract}
The Sun is a main source of high energy neutrinos.
These neutrinos appear as secondary particles after
the Sun absorbs high-energy cosmic rays, 
that find 
there a low-density environment (much thinner
than our atmosphere) where most secondary pions, kaons and muons 
can decay before they lose energy. 
The main uncertainty in a calculation of the solar neutrino flux 
is due to the effects of the magnetic fields on 
the absorption rate of charged cosmic rays.
We use recent data from HAWC on the cosmic-ray shadow of the Sun 
to estimate this rate. We evaluate the 
solar neutrino flux and show that 
at 1 TeV it is over ten times 
larger than the atmospheric one at zenith  
$\theta_z=30^\circ/150^\circ$. 
The flux that we obtain has a 
distinct spectrum and flavor composition: it is harder and richer in antineutrinos
and tau/electron neutrinos than the atmospheric background. This solar flux could
be detected in  current and upcoming
neutrino telescopes.
KM3NeT, in particular, looks very promising:
it will see the Sun high in the sky
(the atmospheric flux is lower there than near the horizon) 
and expects a very good angular resolution (the Sun's radius 
is just $0.27^\circ$). 
\end{abstract}

%\pacs{12.60.Jv, 11.25.Mj, 95.35.+d}% PACS, the Physics and Astronomy
%\keywords{Galactic magnetic fields, TeV cosmic rays}
                              %Use showkeys class option if keyword
\newpage

\section{Introduction}
High-energy astroparticles provide a
picture of the sky that complements the traditional one from light
in different frequencies. They reach the Earth with 
a spectrum that 
extends up to $10^{11}$ GeV, millions of times above 
the energies that we are able to achieve at particle colliders.
Their study for over 100 years has defined a puzzle 
that, although
not complete yet, has helped us to understand the 
environment where these particles are produced: supernovas, 
pulsars, active galactic nuclei or gamma ray bursts, 
where nature reaches its most extreme conditions. 

Neutrinos are an essential piece in that puzzle. They appear
whenever a cosmic proton or a heavier nucleus interacts with 
matter or light and fragments into secondary
neutrons, pions and kaons, which decay giving leptons.
In astronomy neutrinos are a {\it unique} messenger: unlike 
charged cosmic rays (CRs), they are not deflected by magnetic
fields and point to the source; unlike gamma rays, 
they can propagate through a dense medium and reach
the Earth unscattered. Their promise,
however, faces two main challenges. First, being 
weakly-interacting particles, they are very difficult to
detect: neutrino telescopes require large volumes in 
order to register just a few events. In addition, 
neutrinos are  
constantly produced when high-energy cosmic
rays enter the atmosphere and start an air shower.
Any astrophysical signal must then be
separated from this atmospheric background. Despite that, 
the recent discovery by IceCube \cite{Aartsen:2013jdh} 
of a diffuse flux of 
cosmic origin proves that high-energy neutrino astronomy 
is indeed possible.

Here we will discuss an astrophysical  flux
that has a precise location in the sky and a known
spectrum and composition:
the high-energy neutrino flux from our Sun. 
Proposed in the early 90's \cite{Seckel:1991ffa,Moskalenko:1991hm},
this flux has recently attracted renewed attention 
\cite{Arguelles:2017eao,Ng:2017aur,Edsjo:2017kjk}.
Its interest  is 
threefold. First of all, the solar flux
is well above the atmospheric background. 
If detected and characterized, it could 
be used to calibrate the energy and the
angular resolution of neutrino telescopes. 
KM3Net \cite{Adrian-Martinez:2016zzs}, in particular, 
will be able to follow the Sun at relatively vertical
directions ($\theta_z\ge 135^\circ$ or 
$\theta_z\le 45^\circ$):
the high-energy atmospheric background is much smaller
there than from the near-horizontal directions typical
at IceCube.
In addition, the analysis of this neutrino flux would certainly 
bring valuable information about the 
solar magnetic field \cite{Seckel:1991ffa}.
Finally, as emphasized in the most recent work
\cite{Arguelles:2017eao,Ng:2017aur,Edsjo:2017kjk},
this flux is itself 
a strong background in the search for  
dark matter annihilation in the  Sun 
\cite{Adrian-Martinez:2013ayv,Adrian-Martinez:2016gti,
Ardid:2017lry,Aartsen:2016zhm}.

Our calculation will differ from previous ones in some significant
aspects. The results at $E<1$ TeV depend crucially on 
the magnetic fields present in the inner solar 
system \cite{Seckel:1991ffa}. In the next section we argue that 
HAWC data \cite{Enriquez:2015nva}
on the cosmic ray shadow of the Sun can be of use for an
estimate. In our analysis we pay special attention to the 
yields in hadron collisions: we use
EPOS-LHC \cite{Pierog:2013ria} to deduce the (energy dependent)
yields $f_{hh'}(x,E)$ of all the long-lived species 
($p,n,\bar p,\bar n, \pi^\pm, K^\pm,K_L$). In addition,
in pion and kaon decays we will 
distinguish the production of positive
and negative helicity muons, which imply very different 
neutrino yields. Finally, our solar model takes into account
the ionization of the hydrogen and helium in the interior of the Sun, 
implying a lower rate of energy loss as muons propagate.

\section{Magnetic effects on cosmic rays}
The magnetic effects of the Sun on CRs may be separated into 
three basic regions. 

\begin{figure}[!t]
\vspace{-3cm}
\begin{center}
\includegraphics[width=0.6\linewidth]{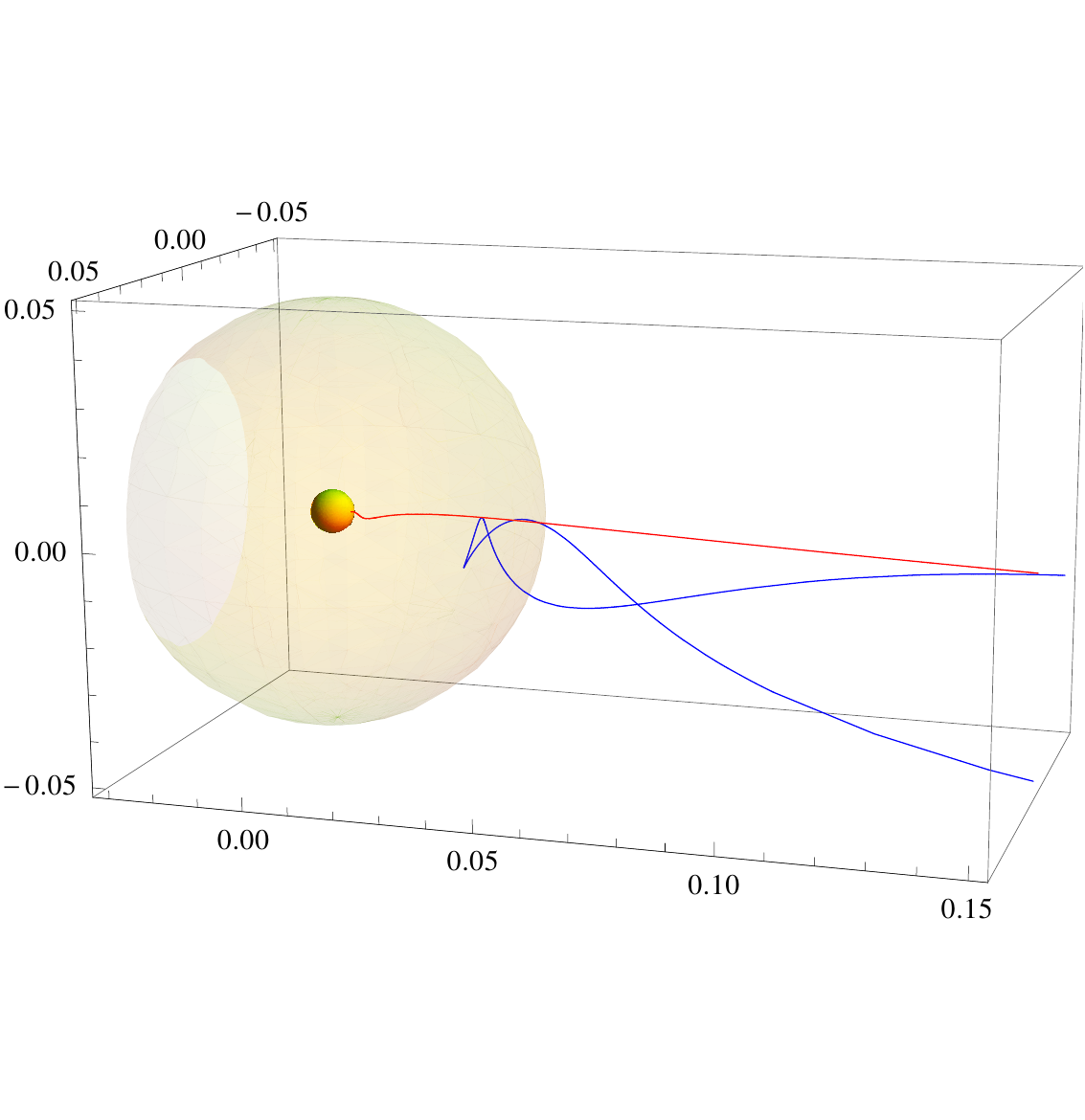}
\end{center}
\vspace{-2cm}
\caption{Trajectories through the Parker field
with origin at the Earth
for 1 TeV (reflected) and 5 TeV (absorbed)
protons. We have shaded the region 
$R< 10 R_\odot$ (in AU).
\label{fig1}}
\end{figure}

{\it (i)}
At distances $R\ge 10 R_\odot$ \cite{Tautz:2010vk} and $E\ge 100$ GeV 
the Parker (interplanetary) field \cite{Parker:1958zz}
will basically define ballistic 
trajectories. Many of these trajectories 
will be magnetically
mirrored before getting to that limit 
(see Fig.~\ref{fig1}), but CRs
can always find a ballistic trajectory close to 
a field line that connects the Earth with the region
$R\approx 10 R_\odot$.

{\it (ii)}
At $R< 10 R_\odot$ the field
lines tend to co rotate with the Sun; magnetic fluctuations become
dominant and the field intensity is very dependent on the
phase in the solar cycle. In the corona, 
at midlatitudes and the equator most field
lines are closed into loops that 
start and finish in the solar 
surface, whereas 
the interplanetary field lines are pushed to the
polar regions. Above the solar transition region
(between the chromosphere and the corona)
the propagation of
$E\le 1$ TeV CRs is probably best described by
a diffusion equation. In particular, 
an energy-dependent diffusion coefficient that 
decreases as CRs approach the Sun's surface
would favor their {\it reflection} 
by the increasing magnetic field strength
that they face.

{\it (iii)}
In the chromosphere and the photosphere 
it is gas pressure and fluid dynamics (instead of the
magnetic field) what dictates the solar structure.
One could use geometric arguments and individual
CR trajectories to estimate the absorption rate and,
most remarkably, to study the possibility 
of an {\it albedo} flux: the flux of cascade
products reflected from the surface 
\cite{Seckel:1991ffa} (see also \cite{Zhou:2016ljf} 
for a discussion of
the gamma-ray flux from the solar disk).

We can use data on the CR shadow of the Sun,
which was first seen by TIBET \cite{amenemori2000} 
and then by other observatories, to justify and 
quantify these magnetic effects. The solar field acts
on CRs as a magnetic lens, 
and we know from 
Liouville theorem that its only possible effect on the 
primary flux is to create a shadow: a
lens (including a mirror) will not make
anisotropic an isotropic flux, 
but it may interrupt (absorb) trajectories 
that were aiming to the Earth. Therefore, the presence
of a shadow reveals the absorption of CRs by the
Sun.

Let us focus on the data taken by 
HAWC \cite{Enriquez:2015nva} 
during the years 2013-2014, 
near a solar maximum. HAWC  
detected the shadow already at $E\approx 2$ TeV. 
It is not a black
disk (a 100\% CR deficit) of $0.27^\circ$ radius (the angular radius of the 
Sun); instead, the shadow is a deficit that decreases radially 
as we move away from
the angular position of the Sun:
\beq
d(\theta)\approx -A \,\exp\left(-{\theta^2 \over 2 \sigma^2}\right)\,,
\eeq
being $A$ and $\sigma$ energy-dependent parameters.
At $E\approx 8$ TeV HAWC finds $A=0.005$ and 
$\sigma=1.4^\circ$; the shadow is {\it diluted} both by the experimental
error and by the solar magnetic field 
into an angular area that corresponds quite closely to the
$R< 10 R_\odot$ region that we assumed dominated by
magnetic turbulence. 
Moreover, integrating to find
the total CR deficit we obtain that it represents a $27\%$ of the
full shadow of the Sun, $D_\odot=-\pi \left(0.27^\circ\right)^2$. This 
indicates that the Sun absorbs just $27\%$ of the 8 TeV
CRs that were headed towards the Earth, while the remaining
$73\%$ were deflected (mirrored) and reached us.

At higher energies, $E\approx 50$ TeV, their fit gives $A=0.013$ and 
$\sigma=1.7^\circ$. The shadow is again diffused into
an angular region ten times larger than the actual size
of the Sun, but now we see the whole total deficit, with no 
reflexion:
\beq
D_\odot^{-1} \int {\rm d} \theta\, 2\pi \theta\, d(\theta) 
 \approx 1\,.
\eeq
At these energies there is a full set of 
CR trajectories that were aiming 
to the Earth through the solar magnetic field and were interrupted 
(absorbed) by the Sun. In contrast, at $E\approx 2$ TeV 
($A=0.0013$ and $\sigma=1.2^\circ$) the observed deficit amounts
to just $6\%$ of the Sun's shadow, implying that up to $94\%$
of CRs were unable to reach the surface.

\begin{figure}[!t]
\vspace{-0.5cm}
\begin{center}
\includegraphics[width=0.5\linewidth]{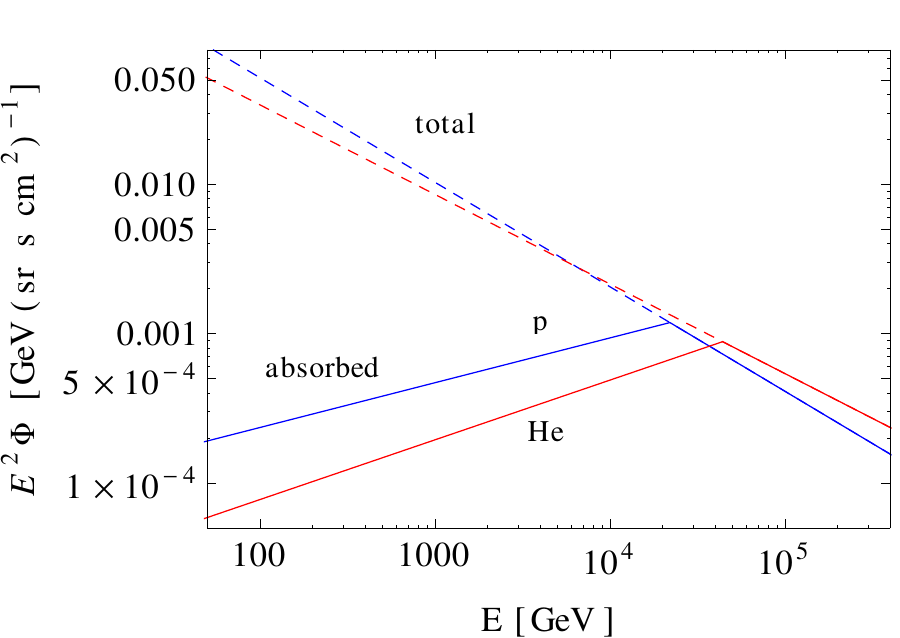}
\end{center}
\vspace{-0.2cm}
\caption{Total and absorbed CR fluxes (see text).
\label{fig2}}
\end{figure}

The previous analysis motivates the following fluxes
(in Fig.~\ref{fig2}).
At $E<E_{\rm knee}$\footnote{We have assumed 
that the proton (He) spectrum changes to a $E^{-3.0}$ 
power law at $10^{6.3}$ ($10^{6.5}$) GeV.} we take \cite{Boezio:2012rr}
a two-component
primary flux with proton and He nuclei 
[in particles/(GeV sr s cm$^2$)]:
\beq
\Phi_p = 1.3 \left( {E\over {\rm GeV}} \right)^{-2.7}
\hspace{1cm}
\Phi_{\rm He} = 0.54 \left( {E\over {\rm GeV}} \right)^{-2.6}
\label{fluxp}
\eeq
At $E>22$ TeV for proton and $E>44$ TeV for helium ({\it i.e.} at
CR rigidities $R>22$ TV) the absorbed
and the total fluxes coincide. At lower energies, however, the spectral
index of the absorbed fluxes changes to $-1.7$ and $-1.6$ for proton 
and He, respectively. 
This change reproduces the deficits 
that we have discussed above.
Since the HAWC data corresponds to a solar maximum, we will also consider
an absorbed flux where the spectral change occurs at lower
energies, $R=22/3$ TV, as a possibility for a quiet Sun.

\section{Solar showers}
In order to understand the solar flux it may be useful to review
the sequence of events taking place after a high-energy 
CR enters the Earth's atmosphere \cite{Gaisser:1990vg}. 
Consider a primary proton of 
$E>100$ GeV, energies where we can neglect the effects of the 
Earth's magnetosphere. Its first interaction
with an air nucleus will typically occur at 20 km of altitude,
after it has crossed a hadronic interaction length
({\it e.g.}, $\lambda_p^{\rm int}=70$ g/cm$^2$ at $E=10^5$ GeV).
As a result, the proton will fragment into a {\it leading baryon} 
carrying 35\% of the initial energy  plus dozens of 
secondary hadrons, mostly mesons. 
The leading baryon
will interact again deeper into the atmosphere, but after 
just four collisions  99\% of its energy will already be 
deposited in the air. Secondary charged pions and kaons, in turn, 
may collide giving more mesons of lower energy or they
may decay giving leptons, {\it e.g.}, $\pi^+ \to \mu^+ \nu_\mu$.
The probability that they do one thing or the other depends on 
their energy and on the air density that they find. For example,
a 10 GeV $\pi^+$ is more likely to decay than collide, since its 
(Lorentz dilated) 560 m decay length is shorter than the typical 
interaction length in the upper 
atmosphere. At 100 GeV, however, the pion will probably hit an 
air nucleus before it has completed its 
5.6 km decay length. Since the higher the energy the less
likely they are to decay, 
the atmospheric neutrino flux from parent pions and kaons
is very suppressed
at high energies, specially from vertical directions 
(from larger zenith angles they face a thinner atmosphere at 
the same slant depth).
Notice also that most muons of energy above 5 GeV will hit
the ground and lose there the energy before they decay 
and give neutrinos, 
$\mu^+ \to e^+ \nu_e \bar\nu_\mu$.

It is easy to realize that the sequence will be quite different
if the same primary proton hits the Sun's surface. 
Let us discuss these differences in some detail.

\begin{figure}[!t]
\vspace{-5cm}
\begin{center}
\includegraphics[width=0.9\linewidth]{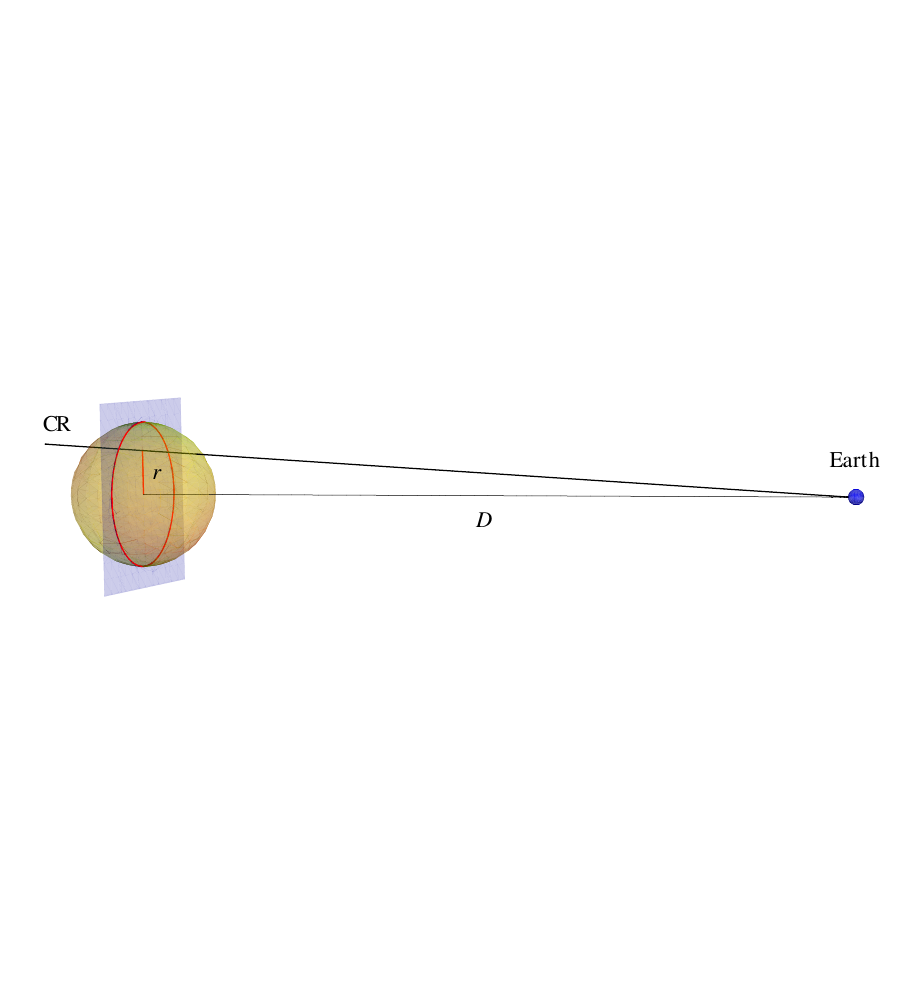}
\end{center}
\vspace{-6cm}
\caption{The CR enters the Sun at a transverse 
distance $r\le R_\odot$, 
producing neutrinos that will reach the Earth at an
angular distance 
$r/D\le 0.27^\circ$ from the Sun's center.
\label{fig3}}
\end{figure}
{\it (i)} 
First of all it will find a 
medium that is much thinner than our atmosphere. The photosphere,
extending
up to 500 km above the Sun's optical surface, has a density between
$3\times 10^{-9}$ and  $2\times 10^{-7}$ g/cm$^3$ (we will use the
solar model in \cite{ChristensenDalsgaard:1996ap}). 
A CR that crosses it vertically will face
a total depth (column density) of just 2.7 g/cm$^2$, whereas
if the CR enters from a radial parameter $r=0.9R_\odot$ (see
Fig.~\ref{fig3}) the total depth of the photosphere increases to 
6.2 g/cm$^2$. Moreover, when the CR goes deeper it will not 
find a sharp
change in the Sun's density. For example,
it takes 1500 km to cross 100 g/cm$^2$ from $r=0$ or up to 2600 km 
from $r=0.9$. The decay length of a 10 TeV
charged pion is 557 km, so most mesons produced there
will have plenty of time to decay and give high-energy neutrinos.
\begin{figure}[!t]
\begin{center}
\includegraphics[width=0.44\linewidth]{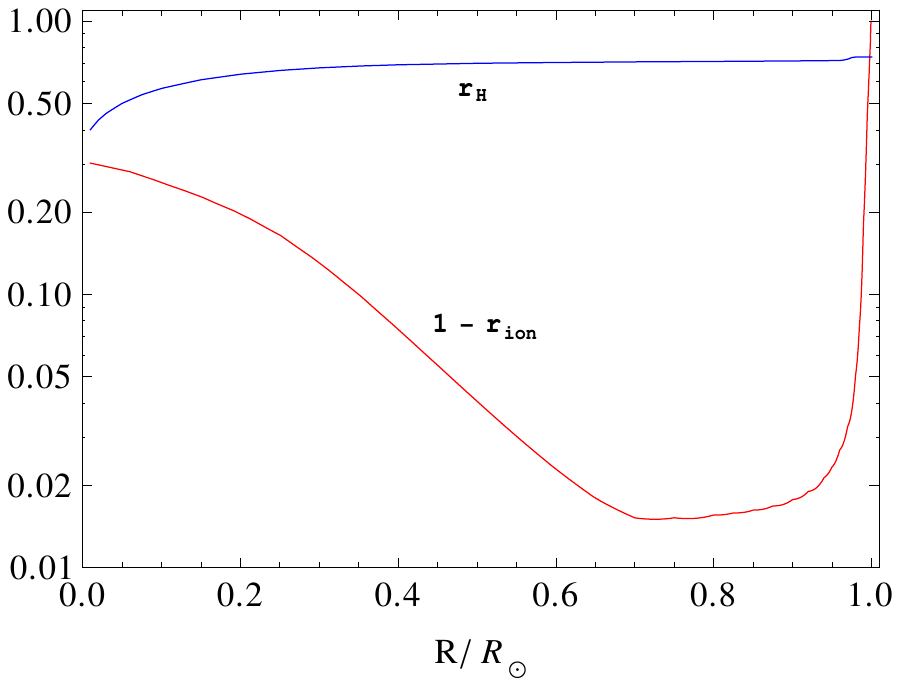}\hspace{0.5cm}
\includegraphics[width=0.47\linewidth]{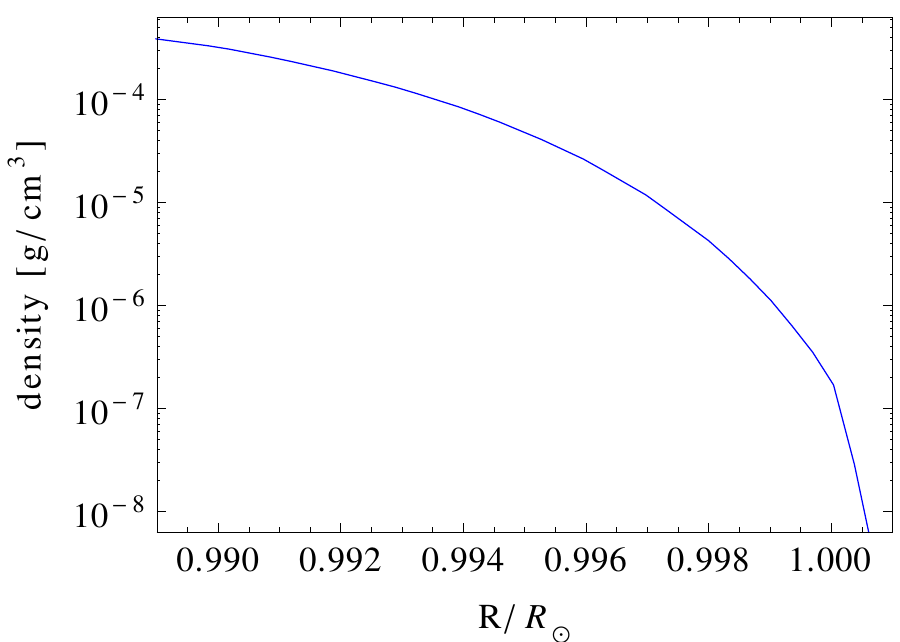}
\end{center}
\vspace{-0cm}
\caption{Fraction of hydrogen ($r_H$) and of ionized matter
($r_{\rm ion}$) at different radii inside the Sun and mass 
density near the surface. 
}
\label{fig4}
\end{figure}

\begin{figure}[!t]
\begin{center}
\includegraphics[width=0.46\linewidth]{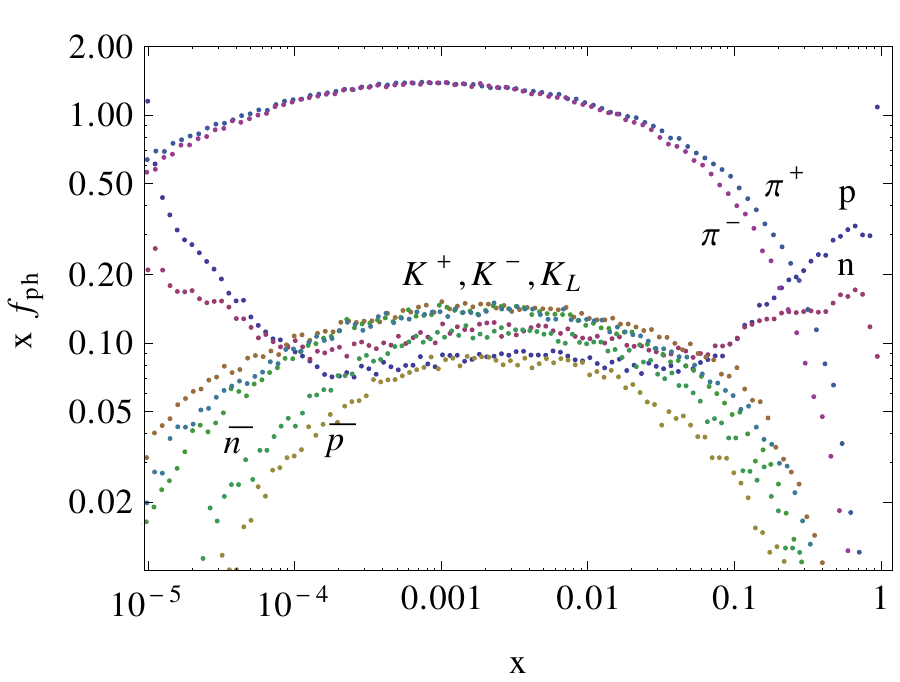}\hspace{0.2cm}
\includegraphics[width=0.46\linewidth]{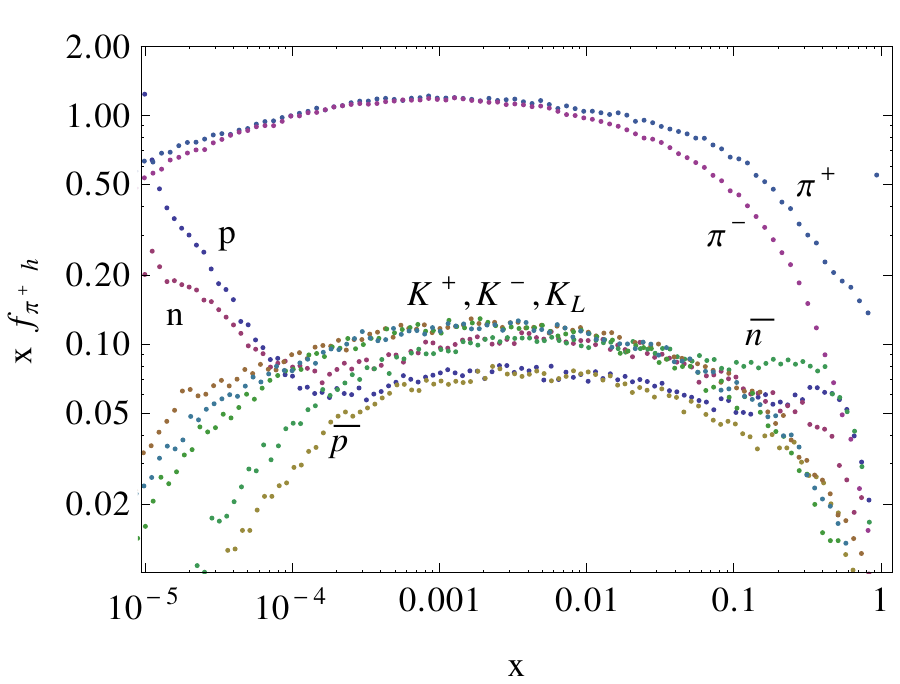}
\end{center}
\vspace{-0cm}
\caption{Yields in $p p$  (left) and  $\pi^+ p$ (right)
collisions at  $E=10^{5}$ GeV. 
}
\label{fig5}
\end{figure}

{\it (ii)} The hadronic shower will then develop within
the initial 2000 km inside the Sun, where the mass density 
($74\%$ hydrogen and the rest mostly $^4$He \cite{Bahcall:2004fg})
is always below $10^{-5}$ g/cm$^3$ (see Fig.~\ref{fig4}). We have
used EPOS-LHC \cite{Pierog:2013ria}
to parametrize the hadronic cross sections 
for nucleon, pion and kaon collisions as well as 
the yields $f_{hh'}(x,E)$ of secondary hadrons 
($h'=p,n,\bar p,\bar n, \pi^\pm, K^\pm,K_L$) carrying
a fraction $x$ of the incident energy $E$ after these collisions.
In particular, we have simulated 50,000 collisions for each
primary and for several energies; we have 
then deduced the yields at those energies
and have used an interpolation to obtain 
$f_{hh'}(x,E)$ in the whole $10$--$10^{8}$ GeV interval.
To illustrate our results, in Fig.~\ref{fig5} 
we plot the yields in $E=10^{5}$ GeV 
$p$ and $\pi^+$ collisions with a proton at rest.

{\it (iii)} After a few thousand km all the hadrons in the shower
have been absorbed or have decayed, and only the neutrino and 
high-energy muon components survive.
The muons will then propagate along the thin medium near the
Sun's surface, so they will have
a significant probability to decay when their energy is still high.
For example, it would take $1.8\times 10^{4}$ km if $r=0$ or 
$3.4\times 10^{4}$ km if $r=0.9R_{\odot}$ to cross a total depth 
of just 15 km w.e.
There are two important factors that change the muon propagation
there relative to the one we observe at the Earth \cite{Olive:2016xmw}. 
First, at high energies energy loss through radiative 
processes will be caused by the low-$Z$ nuclei in the Sun,
and second, these nuclei are partially ionized, resulting into a
reduced rate of muon energy loss also at lower energies.
We estimate a mean energy loss per unit depth $t$ 
\beq
-{{\rm d} \langle E_\mu\rangle \over {\rm d} t} =
\left( 1 - r_{\rm ion}\right) \left[ r_{\rm H} \left( a^{\rm H} - 
a^{\rm He} \right) + a^{\rm He} \right]
+ \left[ r_{\rm H} \left( b^{\rm H} - b^{\rm He} \right) + 
b^{\rm He} \right] E_\mu\,,
\eeq
where $r_{\rm ion}$ is the fraction of ionized matter deduced from
Saha equation (it goes from $10^{-4}$ near the Sun's surface 
to 0.96 at a depth of 15 km w.e., see Fig.~\ref{fig4}), 
$r_{\rm H}$ is the hydrogen fraction,
$a^{\rm H}=4.8$ MeV cm$^{2}$/g, $a^{\rm He}=2.8$ MeV cm$^{2}$/g, 
$b^{\rm H}=2.1\times 10^{-6}$ cm$^{2}$/g, and 
$b^{\rm He}=1.6\times 10^{-6}$ cm$^{2}$/g. 
Notice that 
the higher rate of
muon decays in the Sun relative to terrestrial air showers will 
induce a 
substantially different $\nu_e$:$\nu_\mu$ ratio at $E>10$ GeV.

{\it (iv)} After a depth of 
15 km w.e. most muons have already decayed and we are
left with only neutrinos. These neutrinos, however, still 
have to cross a large
fraction of the Sun's volume before they can emerge from
the opposite side and reach the Earth (see Fig.~\ref{fig3}). 
Their absorption will introduce a significant suppression 
of higher energies
in this $\nu$ flux, specially for 
low values of $r$.
The absorption length $\lambda_\nu$ for a 100 GeV neutrino 
is around 
$7.9\times 10^{12}$ g/cm$^{2}$ \cite{Connolly:2011vc}; 
since the total depth of the Sun goes from 
$3.0\times 10^{12}$ g/cm$^{2}$ at $r=0$ to 
$8.7\times 10^{8}$ g/cm$^{2}$ at $r=0.9R_{\odot}$, most of these 
neutrinos will not be absorbed. 
At $E_\nu=1$ TeV $\lambda_\nu$
 is reduced to $2.8\times 10^{11}$ g/cm$^{2}$, which is
longer than the solar depth only for $r\ge 0.33 R_{\odot}$. At these 
energies the antineutrinos have a 3 times longer absorption length,
which favors them {\it versus} neutrinos. At $E_\nu=100$ TeV we have
$\lambda_\nu=7.6\times 10^{9}$ g/cm$^{2}$, and only the 
neutrinos crossing the Sun
at $r\ge 0.74 R_{\odot}$ will more likely 
emerge than being absorbed.

{\it (v)} A final but important effect are the flavor 
oscillations \cite{Hettlage:1999zr}. As shown in \cite{Fogli:2006jk},
matter effects are suppressed when averaged over the production
region and over the $\nu$ and $\bar \nu$ components, 
and the final flavor is 
dominated by oscillations in vacuum 
between the Sun and the Earth. In particular,
muon and electron neutrinos 
will experience multiple oscillations 
into the $\nu_\tau$ flavor at energies $E<70$ TeV and $E<2$ TeV, 
respectively.
\vspace{0.5cm}

\section{Neutrino fluxes}
As discussed in Section 2, the CRs that we do {\it not} see at the Earth 
have showered near the Sun's surface.
We have also learned in the previous section that 
secondaries of up to several TeV will have plenty of time to decay there
before they collide and lose energy. In contrast with what
happens in the Earth atmosphere, this will be the case for
showers entering the Sun from any zenith angle, and 
independent of how 
{\it curly} the trajectory of secondary 
pions, kaons and muons becomes. We will then assume that the
Sun emission is near isotropic. 
In addition, at  $E< 1$ TeV the absorption of neutrinos
by the Sun will not be significant. 
This means that at these neutrino energies 
any CR trajectory is 
equally good to
determine the $\nu$ flux emitted by the Sun: 
we will take a straight shower 
entering from the opposite side of the Sun like the one 
in Fig.~\ref{fig3}.
Notice that the albedo flux of neutrinos in that side
of the Sun will be compensated by a similar albedo flux in 
the side facing the Earth.

At higher energies ($E\ge 1$ TeV) the neutrino absorption by the
Sun will be important. The pions producing these 
neutrinos, however, are very energetic and their trajectory
will be less affected by the solar magnetic
fields. Therefore, we think
that the straight showers entering the Sun
with parameter $r$ between 0 and $R_\odot$ 
(see Fig.~\ref{fig3}) may provide a good 
approximation for the flux at all neutrino energies.

\begin{figure}[!t]
\begin{center}
\includegraphics[width=0.5\linewidth]{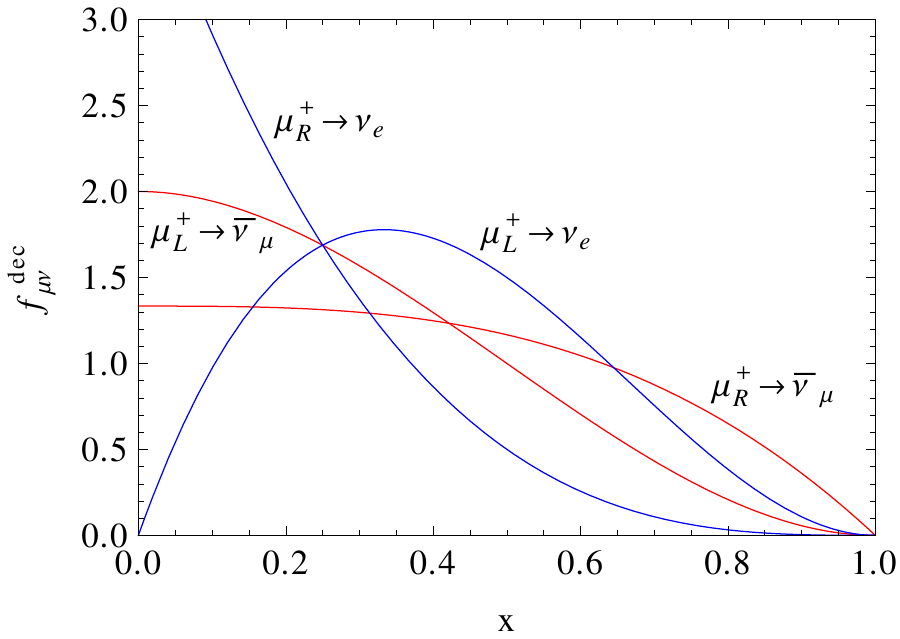}
\end{center}
\vspace{-0.5cm}
\caption{Yields of muon and electron neutrinos in the 
decay of positive ($\mu_R$) and negative ($\mu_L$) helicity
muons. 
}
\label{fig6}
\end{figure}

It is then easy to deduce the transport 
equations for the 9 long-lived hadron species 
($h=p,n,\bar p,\bar n, \pi^\pm, K^\pm,K_L$), muons of left and 
right-handed helicity ($\mu=\mu^\pm_L, \mu^\pm_R$) 
and neutrinos 
($\nu=\nu_{e,\mu}, \bar \nu_{e,\mu}$). 
It is necessary to distinguish between both muon helicities,
as their decay (in Fig.~\ref{fig6}) implies very different 
neutrino distributions 
\cite{Lipari:1993hd}
(see also \cite{Illana:2010gh} for the 
lepton yields from 3-body meson decays). The generic equations are
\beqa
{{\rm d}\Phi_i(E,t)\over {\rm d} t} &=&-\,{\Phi_i(E,t)\over 
\lambda_i^{\rm int}(E,t)} -
{\Phi_i(E,t)\over \lambda_i^{\rm dec}(E,t)} +
\sum_{j=h} \int_0^1 {\rm d}x\; {f_{ji}(x,E/x)\over x}\;
{\Phi_{j}(E/x,t)\over \lambda_{j}^{\rm int}(E/x,t)} +\nonumber\\
&&\sum_{k=h,\mu} \int_0^1 {\rm d}x\; {f_{ki}^{\rm dec}(x,E/x)\over x}\;
{\Phi_{k}(E/x,t)\over \lambda_{j}^{\rm dec}(E/x,t)}\,,
\eeqa
where the sources include both collisions and decays and the
interaction/decay lengths are expressed in g/cm$^2$.
The equations for the muon fluxes have the extra term  
\beq
{{\rm d}\Phi_\mu(E,t)\over {\rm d}t} \supset 
-\,{{\rm d}\langle E_\mu\rangle \over {\rm d} t} \,
{{\rm d}\Phi_\mu(E,t)\over {\rm d} E}
- \Phi_\mu(E,t)\, {{\rm d}\over {\rm d}E_\mu}
\left( {{\rm d} \langle E_\mu\rangle \over {\rm d} t} \right)\,,
\eeq
describing energy loss. We do not include flavor oscillations
in the transport of neutrinos inside the Sun.
The numerical resolution of
these 17 equations provides then the  following results. 

\begin{figure}[!t]
\begin{center}
\includegraphics[width=0.45\linewidth]{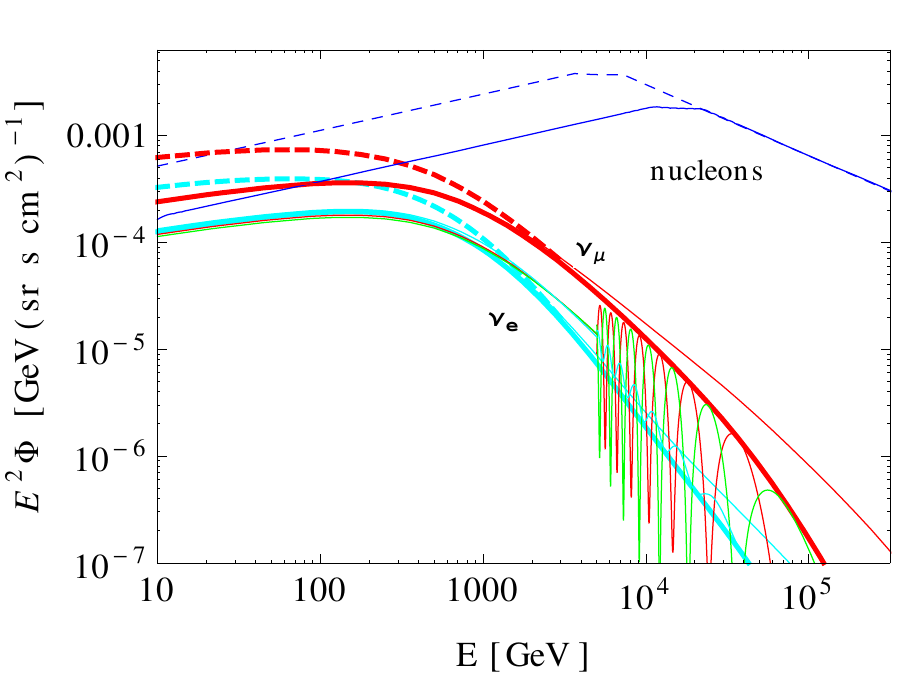}\hspace{0.3cm}
\includegraphics[width=0.45\linewidth]{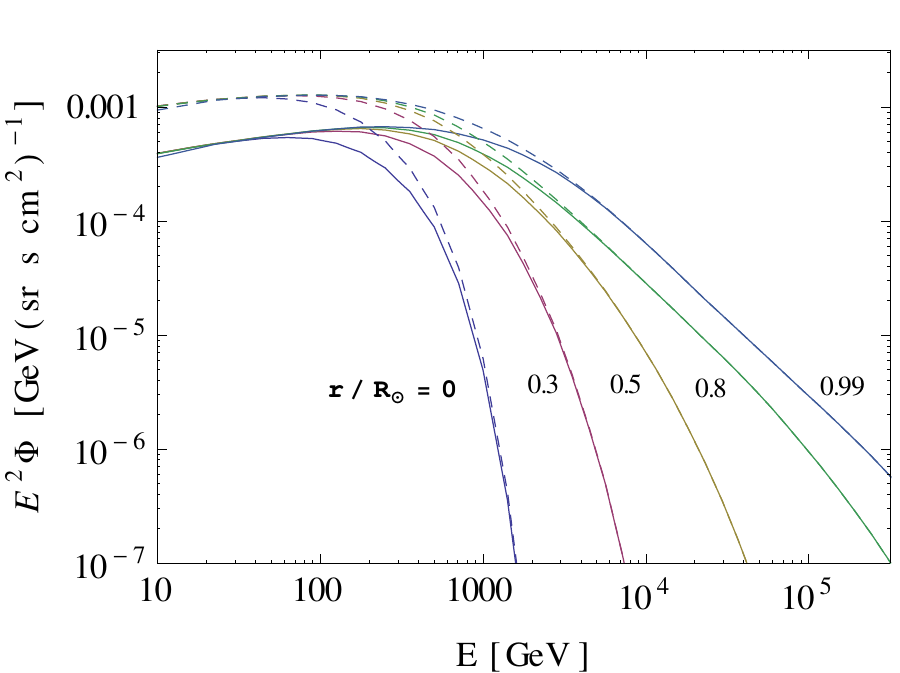}
\end{center}
\vspace{-0cm}
\caption{{\bf Left}: $\nu$+$\bar \nu$
and primary nucleon fluxes for 
$r=2/3\, R_{\odot}$. Red, cyan and green lines correspond to
the $\nu_\mu$, $\nu_e$ and $\nu_\tau$ fluxes, respectively. 
They are given at 15 km w.e. (upper
thin lines), after the partial absorption by the Sun
(thick lines), and at the Earth (lower thin lines).
Below 5 TeV we plot the
averaged oscillations (the lines
for the three flavors almost coincide). {\bf Right}: Total
flux at the Earth for different values of $r$. Solid and dashed
lines correspond to our modelization of the absorption rate for maximum and
minimum of Solar activity, respectively.}
\label{fig7}
\end{figure}

In Fig.~\ref{fig7}--left we take 
$r=2/3\, R_{\odot}$, which is the average
radial distance in a circle, and 
plot {\it (i)} the neutrino flux at a Solar depth of 
15 km w.e.
(thin red and cyan lines for $\nu_\mu$ and $\nu_e$, respectively), 
{\it (ii)} the flux after crossing the Sun (thick
red and cyan lines),
and {\it (iii)} the final $\nu$ flux at the Earth for the three neutrino
flavors (thin cyan, red and green lines for $\nu_e$, $\nu_\mu$ and
$\nu_\tau$, respectively).
In the plot we include 
the primary all-nucleon flux (protons plus 
neutrons bound in helium) for the absorbed flux deduced 
from HAWC near a solar maximum (solid) and another one 
that could correspond to a quieter solar phase (dashes). 
We can see that $\nu$ oscillations play a crucial role: 
at $E\approx 70$ TeV most muon neutrinos have oscillated 
into the $\nu_\tau$ flavor; at
$E\le 1$ TeV the three (averaged) flavors are almost 
undistinguishable, and they coincide with the $\nu_e$
flux before oscillations (thick cyan line). 
In our analysis we have taken the
neutrino masses and mixings deduced in \cite{Gonzalez-Garcia:2014bfa}.

Fig.~\ref{fig7}--right gives the total neutrino flux 
reaching the Earth  from different radial
distances $r$. As expected, at $E>1$ TeV the flux from 
the center ($r=0$) is much
weaker than from the peripheral regions. Notice also that the
lower-energy fluxes do not depend on $r$, indicating that 
the longitudinal development of the shower is independent
of the zenith angle.

\begin{figure}[!t]
\begin{center}
\includegraphics[width=0.46\linewidth]{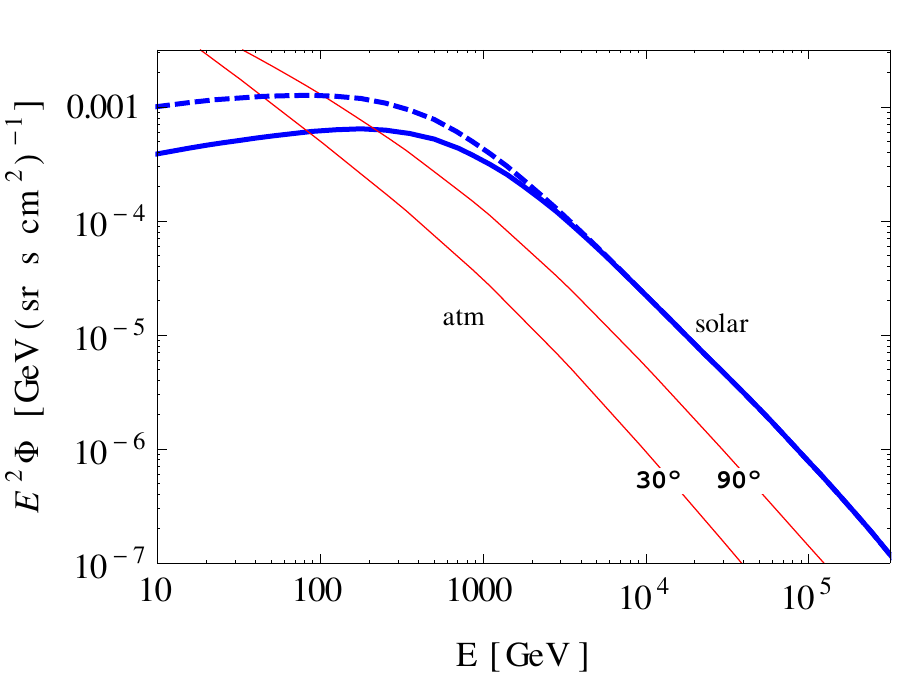}
\end{center}
\vspace{-0cm}
\caption{Comparison of the solar and the atmospheric (for two
zenith angles) fluxes. The first one has been averaged over the whole
solid angle occupied by the Sun 
(a circle of $0.27^\circ$ radius).  Solid and dashed
lines correspond to our modelization of the absorption rate for maximum and
minimum of Solar activity, respectively.
}
\label{fig8}
\end{figure}

Finally, in Fig.~\ref{fig8} we compare the average solar flux
with the atmospheric one \cite{Lipari:1993hd} from two 
different zenith angles.
At 500 GeV 
the solar flux (neutrinos plus
antineutrinos of all flavors) is 7.0 (2.0) times larger than
the atmospheric one
from $\theta_z=30^\circ/150^\circ$ ($\theta_z=90^\circ$).
At 5 TeV, the relative 
difference with the atmospheric flux increases: 
20.5 times larger from $\theta_z=30^\circ/150^\circ$ 
and 4.0 times larger 
from horizontal directions. 
If we restrict to the $\nu_\mu$ flavor, at 5 TeV
the average flux from the Sun is 7.0 and 1.4 times larger
than the atmospheric one from those two inclinations, respectively.
Notice that
the atmospheric background is significantly stronger 
when the Sun is seen at large zenith angles.

Our calculation of the high-energy solar neutrino flux has several 
sources of uncertainty. Different hadronic models 
imply lepton yields that typically differ in a $10\%$ (see the comparison
between EPOS-LHC and SIBYLL \cite{Ahn:2009wx} in \cite{Carceller:2016upo}).
However, at $E_\nu >1$ TeV the main uncertainty in our result
comes from the $10$--$1000$ TeV all-nucleon flux, whose
accuracy may be estimated at $20\%$ \cite{Boezio:2012rr,Olive:2016xmw}. 
At lower neutrino energies our calculation 
relies strongly on the absorption 
rate by the Sun of $0.1$--$10$ TeV CRs, something that is given by HAWC
with a $\approx 20\%$ uncertainty and that is expected to 
change during the solar cycle. The available data from 
HAWC corresponds to a solar maximum in 2013--2014, and
thus the $\approx 100$ GeV $\nu$ flux could increase substantially  
during the next years (a 50\% in our estimate, see Fig.~\ref{fig8}).

\section{Summary and discussion}
It seems quite plausible that the Sun is the most luminous object 
in the sky also for high energy neutrinos.
We think that the existence of a known 
source that gives a signal above the atmospheric
background can be useful for the development of neutrino
telescopes. The atmospheric 
background is certainly stronger from the 
near horizontal directions that point to the Sun  
at IceCube than from the more vertical ones at KM3NeT,
but features like  the presence of a
strong $\nu_\tau$ component (which is absent in the 
atmospheric flux) offer hope that it can be observed also there.

We have correlated this high-energy neutrino signal 
with HAWC's observations of
a CR deficit from the Sun. Our results are qualitatively very
similar to the ones in \cite{Seckel:1991ffa}, but at high
energies imply a neutrino
flux slightly higher than the recent ones obtained
in \cite{Arguelles:2017eao,Ng:2017aur,Edsjo:2017kjk}.
Indeed, 
a measurement of the Sun's shadow 
during its whole 11-year cycle would imply a more precise
prediction of this solar neutrino flux, specially at $E_\nu\le1$ TeV.
Its detection and analysis at neutrino telescopes
will improve our understanding of the magnetic properties
and the internal structure of the Sun.

\section*{Acknowledgments}
The author would like to thank Miquel Ardid, Eduardo Battaner, 
John Beacom, Joaqu\'\i n Castellano, 
Joakim Edsjo, Juanjo Hern\'andez, Kenny Ng,
Aaron Vincent and Juande Zornoza for comments and discussions.
This work has been supported by MICINN of Spain 
(FPA2013-47836, FPA2015-68783-REDT, FPA2016-78220, 
Consolider-Ingenio {\bf MultiDark} CSD2009-00064) 
and by Junta de Andaluc\'\i a (FQM101).


\begin{thebibliography}{99} 

\bibitem{Aartsen:2013jdh}
  M.~G.~Aartsen {\it et al.} [IceCube Collaboration],
  %``Evidence for High-Energy Extraterrestrial Neutrinos at the IceCube Detector,''
  Science {\bf 342} (2013) 1242856.
%  doi:10.1126/science.1242856
%  [arXiv:1311.5238 [astro-ph.HE]].
  %%CITATION = doi:10.1126/science.1242856;%%
  %524 citations counted in INSPIRE as of 14 Feb 2017

\bibitem{Seckel:1991ffa}
  D.~Seckel, T.~Stanev and T.~K.~Gaisser,
  %``Signatures of cosmic-ray interactions on the solar surface,''
  Astrophys.\ J.\  {\bf 382} (1991) 652.
%  doi:10.1086/170753
  %%CITATION = doi:10.1086/170753;%%
  %66 citations counted in INSPIRE as of 14 Jun 2017

\bibitem{Moskalenko:1991hm}
  I.~V.~Moskalenko, S.~Karakula and W.~Tkaczyk,
  %``The Sun as the source of VHE neutrinos,''
  Astron.\ Astrophys.\  {\bf 248} (1991) L5.
  %%CITATION = AAEJA,248,L5;%%
  %6 citations counted in INSPIRE as of 14 Jun 2017

\bibitem{Arguelles:2017eao}
  C.~A.~Argüelles, G.~de Wasseige, A.~Fedynitch and B.~J.~P.~Jones,
  ``Solar Atmospheric Neutrinos and the Sensitivity Floor for Solar Dark Matter Annihilation Searches,''
  arXiv:1703.07798 [astro-ph.HE].
  %%CITATION = ARXIV:1703.07798;%%
  %2 citations counted in INSPIRE as of 14 Jun 2017

\bibitem{Ng:2017aur}
  K.~C.~Y.~Ng, J.~F.~Beacom, A.~H.~G.~Peter and C.~Rott,
  ``Solar Atmospheric Neutrinos: A New Neutrino Floor for 
Dark Matter Searches,''
  arXiv:1703.10280 [astro-ph.HE].
  %%CITATION = ARXIV:1703.10280;%%
  %1 citations counted in INSPIRE as of 14 Jun 2017

\bibitem{Edsjo:2017kjk}
  J.~Edsjo, J.~Elevant, R.~Enberg and C.~Niblaeus,
  ``Neutrinos from cosmic ray interactions in the Sun,''
  arXiv:1704.02892 [astro-ph.HE].
  %%CITATION = ARXIV:1704.02892;%%

\bibitem{Adrian-Martinez:2016zzs}
  S.~Adri\'an-Mart\'\i nez {\it et al.} [KM3NeT Collaboration],
  %``Intrinsic limits on resolutions in muon- and electron-neutrino charged-current events in the KM3NeT/ORCA detector,''
  JHEP {\bf 1705} (2017) 008
%  doi:10.1007/JHEP05(2017)008
%  [arXiv:1612.05621 [physics.ins-det]].
  %%CITATION = doi:10.1007/JHEP05(2017)008;%%
  %1 citations counted in INSPIRE as of 03 Jun 2017


\bibitem{Adrian-Martinez:2013ayv}
  S.~Adrian-Martinez {\it et al.} [ANTARES Collaboration],
  %``First results on dark matter annihilation in the Sun using the ANTARES neutrino telescope,''
  JCAP {\bf 1311} (2013) 032.
%  doi:10.1088/1475-7516/2013/11/032
%  [arXiv:1302.6516 [astro-ph.HE]].
  %%CITATION = doi:10.1088/1475-7516/2013/11/032;%%
  %73 citations counted in INSPIRE as of 02 Jun 2017

\bibitem{Adrian-Martinez:2016gti}
  S.~Adrian-Martinez {\it et al.} [ANTARES Collaboration],
  %``Limits on Dark Matter Annihilation in the Sun using the ANTARES Neutrino Telescope,''
  Phys.\ Lett.\ B {\bf 759} (2016) 69.
%  doi:10.1016/j.physletb.2016.05.019
%  [arXiv:1603.02228 [astro-ph.HE]].
  %%CITATION = doi:10.1016/j.physletb.2016.05.019;%%
  %25 citations counted in INSPIRE as of 14 Jun 2017

\bibitem{Ardid:2017lry}
  M.~Ardid, I.~Felis, A.~Herrero and J.~A.~Mart\'\i nez-Mora,
  %``Constraining Secluded Dark Matter models with the public data from the 79-string IceCube search for dark matter in the Sun,''
  JCAP {\bf 1704} (2017) no.04,  010.
%  doi:10.1088/1475-7516/2017/04/010
%  [arXiv:1701.08863 [astro-ph.HE]].
  %%CITATION = doi:10.1088/1475-7516/2017/04/010;%%
  %3 citations counted in INSPIRE as of 05 Jun 2017

\bibitem{Aartsen:2016zhm}
  M.~G.~Aartsen {\it et al.} [IceCube Collaboration],
  %``Search for annihilating dark matter in the Sun with 3 years of IceCube data,''
  Eur.\ Phys.\ J.\ C {\bf 77} (2017) no.3,  146.
%  doi:10.1140/epjc/s10052-017-4689-9
%  [arXiv:1612.05949 [astro-ph.HE]].
  %%CITATION = doi:10.1140/epjc/s10052-017-4689-9;%%
  %22 citations counted in INSPIRE as of 14 Jun 2017

\bibitem{Enriquez:2015nva}
  O.~Enriquez-Rivera {\it et al.} [HAWC Collaboration],
  %``The Galactic cosmic-ray Sun shadow observed by HAWC,''
  PoS ICRC {\bf 2015} (2016) 099.
 % [arXiv:1508.07351 [astro-ph.SR]].
  %%CITATION = ARXIV:1508.07351;%%
  %2 citations counted in INSPIRE as of 02 Jun 2017

\bibitem{Pierog:2013ria}
  T.~Pierog, I.~Karpenko, J.~M.~Katzy, E.~Yatsenko and K.~Werner,
  %``EPOS LHC: Test of collective hadronization with data measured at the CERN Large Hadron Collider,''
  Phys.\ Rev.\ C {\bf 92} (2015) 034906.
%  doi:10.1103/PhysRevC.92.034906
%  [arXiv:1306.0121 [hep-ph]].
  %%CITATION = doi:10.1103/PhysRevC.92.034906;%%
  %183 citations counted in INSPIRE as of 30 Sep 2016

\bibitem{Tautz:2010vk}
  R.~C.~Tautz, A.~Shalchi and A.~Dosch,
  %``Simulating Heliospheric and Solar Particle Diffusion using the Parker Spiral Geometry,''
  J.\ Geophys.\ Res.\ Space Phys.\  {\bf 116} (2011) 2102.
%  doi:10.1029/2010JA015936
%  [arXiv:1011.3325 [astro-ph.HE]].
  %%CITATION = doi:10.1029/2010JA015936;%%

\bibitem{Parker:1958zz}
  E.~N.~Parker,
  %``Dynamics of the Interplanetary Gas and Magnetic Fields,''
  Astrophys.\ J.\  {\bf 128} (1958) 664.
%  doi:10.1086/146579
  %%CITATION = doi:10.1086/146579;%%
  %304 citations counted in INSPIRE as of 02 Jun 2017

\bibitem{Zhou:2016ljf} B.~Zhou, K.~C.~Y.~Ng, J.~F.~Beacom and A.~H.~G.~Peter, ``TeV Solar Gamma Rays From Cosmic-Ray Interactions,''
arXiv:1612.02420 [astro-ph.HE].
%%CITATION = ARXIV:1612.02420;%%
%4 citations counted in INSPIRE as of 17 Jun 2017 

\bibitem{amenemori2000} 
M.~Amenemori {\it et al.} [TIBET AS$\gamma$ Collaboration],  
Astrophys.\ J.\  {\bf 541} (2000) 1051.

\bibitem{Boezio:2012rr}
  M.~Boezio and E.~Mocchiutti,
  %``Chemical Composition of Galactic Cosmic Rays with Space Experiments,''
  Astropart.\ Phys.\  {\bf 39-40} (2012) 95.
%  doi:10.1016/j.astropartphys.2012.05.007
%  [arXiv:1208.1406 [hep-ex]].
  %%CITATION = doi:10.1016/j.astropartphys.2012.05.007;%%

\bibitem{Gaisser:1990vg}
  T.~K.~Gaisser,
  %``Cosmic rays and particle physics,''
  Cambridge, UK: Univ. Pr. (1990) 279 p
  %155 citations counted in INSPIRE as of 02 Jun 2017

\bibitem{ChristensenDalsgaard:1996ap}
  J.~Christensen-Dalsgaard {\it et al.},
  %``The current state of solar modeling,''
  Science {\bf 272} (1996) 1286.
%  doi:10.1126/science.272.5266.1286
  %%CITATION = doi:10.1126/science.272.5266.1286;%%
  %309 citations counted in INSPIRE as of 30 May 2017

\bibitem{Bahcall:2004fg}
  J.~N.~Bahcall and M.~H.~Pinsonneault,
  %``What do we (not) know theoretically about solar neutrino fluxes?,''
  Phys.\ Rev.\ Lett.\  {\bf 92} (2004) 121301.
%  doi:10.1103/PhysRevLett.92.121301
%  [astro-ph/0402114].
  %%CITATION = doi:10.1103/PhysRevLett.92.121301;%%
  %289 citations counted in INSPIRE as of 31 May 2017

\bibitem{Olive:2016xmw}
  C.~Patrignani {\it et al.} [Particle Data Group],
  %``Review of Particle Physics,''
  Chin.\ Phys.\ C {\bf 40} (2016) no.10,  100001.
%  doi:10.1088/1674-1137/40/10/100001
  %%CITATION = doi:10.1088/1674-1137/40/10/100001;%%
  %934 citations counted in INSPIRE as of 04 Jun 2017

\bibitem{Connolly:2011vc}
  A.~Connolly, R.~S.~Thorne and D.~Waters,
  %``Calculation of High Energy Neutrino-Nucleon Cross Sections and Uncertainties Using the MSTW Parton Distribution Functions and Implications for Future Experiments,''
  Phys.\ Rev.\ D {\bf 83} (2011) 113009.
%  doi:10.1103/PhysRevD.83.113009
%  [arXiv:1102.0691 [hep-ph]].
  %%CITATION = doi:10.1103/PhysRevD.83.113009;%%
  %66 citations counted in INSPIRE as of 01 Jun 2017

\bibitem{Hettlage:1999zr}
  C.~Hettlage, K.~Mannheim and J.~G.~Learned,
  %``The Sun as a high-energy neutrino source,''
  Astropart.\ Phys.\  {\bf 13} (2000) 45.
%  doi:10.1016/S0927-6505(99)00120-6
%  [astro-ph/9910208].
  %%CITATION = doi:10.1016/S0927-6505(99)00120-6;%%
  %30 citations counted in INSPIRE as of 20 Jun 2017

\bibitem{Fogli:2006jk}
  G.~L.~Fogli, E.~Lisi, A.~Mirizzi, D.~Montanino and P.~D.~Serpico,
  %``Oscillations of solar atmosphere neutrinos,''
  Phys.\ Rev.\ D {\bf 74} (2006) 093004.
%  doi:10.1103/PhysRevD.74.093004
%  [hep-ph/0608321].
  %%CITATION = doi:10.1103/PhysRevD.74.093004;%%
  %34 citations counted in INSPIRE as of 20 Jun 2017

\bibitem{Lipari:1993hd}
  P.~Lipari,
  %``Lepton spectra in the earth's atmosphere,''
  Astropart.\ Phys.\  {\bf 1} (1993) 195.
%  doi:10.1016/0927-6505(93)90022-6
  %%CITATION = doi:10.1016/0927-6505(93)90022-6;%%
  %371 citations counted in INSPIRE as of 05 Oct 2016

\bibitem{Illana:2010gh}
  J.~I.~Illana, P.~Lipari, M.~Masip and D.~Meloni,
  %``Atmospheric lepton fluxes at very high energy,''
  Astropart.\ Phys.\  {\bf 34} (2011) 663.
%  doi:10.1016/j.astropartphys.2011.01.001
%  [arXiv:1010.5084 [astro-ph.HE]].
  %%CITATION = doi:10.1016/j.astropartphys.2011.01.001;%%
  %25 citations counted in INSPIRE as of 01 Jun 2017

\bibitem{Gonzalez-Garcia:2014bfa}
  M.~C.~Gonzalez-Garcia, M.~Maltoni and T.~Schwetz,
  %``Updated fit to three neutrino mixing: status of leptonic CP violation,''
  JHEP {\bf 1411} (2014) 052.
%  doi:10.1007/JHEP11(2014)052
%  [arXiv:1409.5439 [hep-ph]].
  %%CITATION = doi:10.1007/JHEP11(2014)052;%%
  %351 citations counted in INSPIRE as of 13 Jul 2016

\bibitem{Ahn:2009wx}
  E.~J.~Ahn, R.~Engel, T.~K.~Gaisser, P.~Lipari and T.~Stanev,
  %``Cosmic ray interaction event generator SIBYLL 2.1,''
  Phys.\ Rev.\ D {\bf 80} (2009) 094003.
%  doi:10.1103/PhysRevD.80.094003
 % [arXiv:0906.4113 [hep-ph]].
  %%CITATION = doi:10.1103/PhysRevD.80.094003;%%
  %361 citations counted in INSPIRE as of 22 Sep 2017

\bibitem{Carceller:2016upo}
  J.~M.~Carceller and M.~Masip,
  %``Diffuse flux of galactic neutrinos and gamma rays,''
  JCAP {\bf 1703} (2017)   013.
%  doi:10.1088/1475-7516/2017/03/013
%  [arXiv:1610.02552 [astro-ph.HE]].
  %%CITATION = doi:10.1088/1475-7516/2017/03/013;%%
  %1 citations counted in INSPIRE as of 22 Sep 2017





\end{thebibliography}
\end{document}